\documentclass[aps, prl, onecolumn, preprint, endfloats*]{revtex4}

\usepackage{graphicx}
\usepackage{amssymb,amsmath, wasysym, lineno}

\usepackage{xr}
\usepackage{natbib}
\bibpunct{}{}{,}{s}{}{} 

\newcommand{\nm}[1]{\ensuremath{#1\ \mathrm{nm}}}
\newcommand{\um}[1]{\ensuremath{#1\ \mathrm{\mu m}}}

\newcommand{\uW}[1]{\ensuremath{#1\ \mathrm{\mu W}}}

\begin{document}
\title{Adaptive control of necklace states in a photonic crystal waveguide}

\author{Emre Y\"uce,$^{1,2,\ast}$ Jin Lian,$^{1}$ Sergei Sokolov,$^{1,5}$ Jacopo Bertolotti,$^{3}$ Sylvian Combri\'e,$^{4}$  Alfredo De Rossi,$^{4}$ Ga\"elle Lehoucq,$^{4}$ and Allard P. Mosk$^{1,5}$}
\affiliation{
$^{1}$Complex Photonic Systems (COPS), MESA+ Institute for Nanotechnology, University of Twente, P.O. Box 217, 7500 AE Enschede, The Netherlands \\
$^{2}$ The Center for Solar Energy Research and Applications (G\"UNAM), Department of Physics, Middle East Technical University, 06800 Ankara, Turkey \\
$^{3}$Physics and Astronomy Department, University of Exeter, Stocker Road, Exeter EX4 4QL,UK\\
$^{4}$Thales Research and Technology, Route Départementale 128, \\91767 Palaiseau, France\\
$^{5}$Debye Institute for Nanomaterials Science, University of Utrecht, PO Box 80000, 3508 TA Utrecht, The Netherlands.\\
}
\date{\today}

\maketitle


{\bfseries
Resonant cavities with high quality factor and small mode volume provide crucial enhancement of light-matter interactions in nanophotonic devices that transport and process classical and quantum information. 
The production of functional circuits containing many such cavities remains a major challenge as inevitable imperfections in the fabrication detune the cavities, which strongly affects functionality such as transmission. 
In photonic crystal waveguides, intrinsic disorder gives rise to high-Q localized resonances through Anderson localization, however their location and resonance frequencies are completely random, which hampers functionality.
We present an adaptive holographic method to gain reversible control on these randomly localized modes by locally modifying the refractive index.
We show that our method can dynamically form or break highly transmitting necklace states, which is an essential step towards photonic-crystal based quantum networks and signal processing circuits.
}

Disorder-induced scattering of light is commonly regarded as a loss mechanism as it degrades transmission~\cite{koenderink.2005.prb, huhes.2005.prl, Brosi.2008.oe, sokolov.2017.oe}.
However, scattering can also give rise to localized resonant modes,  with a potentially very high quality factor.  
In particular, close to the band-edge of a photonic crystal membrane, the formation of Anderson localized modes~\cite{anderson.1958.pr} is a natural consequence of intrinsic disorder~\cite{koenderink.2005.prb,huhes.2005.prl,Topolancik.2007.prl, lodahl.2010.science}.
The confinement strength, lifetime, and the spatial profile of these modes are statistically controlled by the dimensionality of the system and the strength of the random scattering ~\cite{faggiani.2016.scirep}. 
In an irreversible manner, light-induced oxidation of the surface was used to control such a single localized mode~\cite{riboli.2014.natmat}.
To date, no reversible control of  localized resonant modes has been demonstrated, which is an essential step towards disorder-resistant programmable photonic circuits.

Here, we develop a novel approach, where the high-Q modes generated by the structural randomness are reversibly controlled and coupled into a high-transmission necklace state~\cite{pendry.1987.JPC}.
We first all-optically locate the relevant Anderson localized modes and subsequently tune their resonance wavelengths independently. 
By  identifying and tuning a link mode, we can form or break necklace states and thereby program the transmitted signal.

\begin {figure}[ht]
\begin {center}
\includegraphics [width=0.6\columnwidth ]{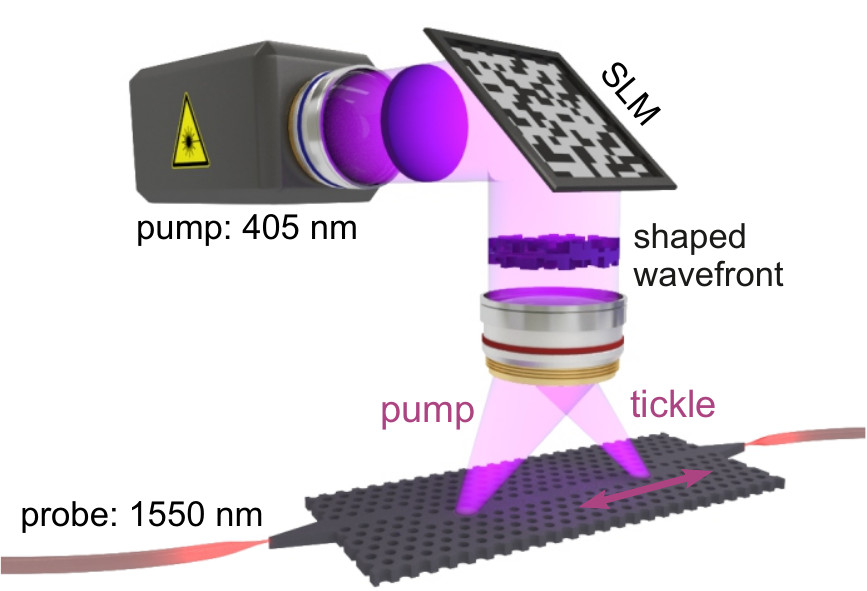}
\caption {\emph { Schematic of the experiment. A narrowband signal beam is coupled to the GaInP photonic crystal waveguide via a tapered fiber.
A second tapered fiber is positioned at the other end of the sample to guide the transmitted signal.
The pump beam (violet) incides normal to the photonic crystal plane and is spatially structured by a spatial light modulator (SLM) such that multiple spots with different powers are formed on the sample. 
}}
\label {Fig1}
\end {center}
\end {figure}

A versatile apparatus is used to control the localized modes in our photonic crystal waveguide~\cite{sokolov.2015.apl}. 
The setup is shown in Fig.~\ref{Fig1} and consists of a tunable laser at the conventional (C) telecom band and a \nm{405} diode laser that are the sources of the signal and pump light, respectively. 
The resonance wavelengths of the localized modes are determined from the transmission spectrum. 
A spatially structured pump beam is generated by a spatial light modulator (SLM) that projects digital holograms. 
This holographic control of the pump beam enables us to project multiple independently controlled pump spots on the sample, which we use to locate, tune, and perturb the localized modes.
Here, we study the mode profiles using a novel ``pump-tickle-probe'' strategy, where a first strong pump beam is used to perturb the modes, and a weak secondary pump beam we designate as ``tickle'' is used to elucidate the spatial profile of the perturbed modes while inducing negligible further perturbation.
The weak 405-nm ``tickle''  beam (\uW{16}) introduces a local thermo-optic perturbation ($\delta n \approx 10^{-4}$) that effectively shifts the resonance of any mode it spatially overlaps with by up to $\approx \nm{0.4}$. 
Measuring the resonance wavelength shift as a function of the tickle beam position, we infer the position of the localized modes~\cite{lian.2016.optexp}.
Subsequently, we use the strong primary pump beam to tune a targeted mode.
In our measurements, the typical pump power is in the order of  \uW{100}, which induces a local index change as large as $\delta n \approx 10^{-3}$ through thermo-optic perturbation. 
The pump beam is kept on the linking modes during all the measurements while we scan the tickle beam and collect transmitted information through the probe beam. 
The probe laser light was coupled to the PhC waveguide using a polarization
maintaining lensed fiber with NA of 0.55. 
We perform our measurements in a flushed $N_2$ environment to reduce oxidation which would otherwise result in irreversible changes~\cite{sokolov.2015.apl}.
 
Our photonic crystal sample is a GaInP membrane structure with a membrane thickness of \nm{180} and the lattice constant of our photonic crystal waveguide is $a=\nm{485}$.
The width of the main (barrier) waveguide is $W_0=0.98\sqrt{3a}$ and its  length is $L=106\,a$.
The main waveguide is side coupled to two access waveguides with widths  $W_1=1.1\sqrt{3a}$ that are positioned at the input and the output facet and serve to couple light in and out of the structure, see Fig.~\ref{Fig1}.

\begin {figure*}[htb]
\begin {center}
\includegraphics {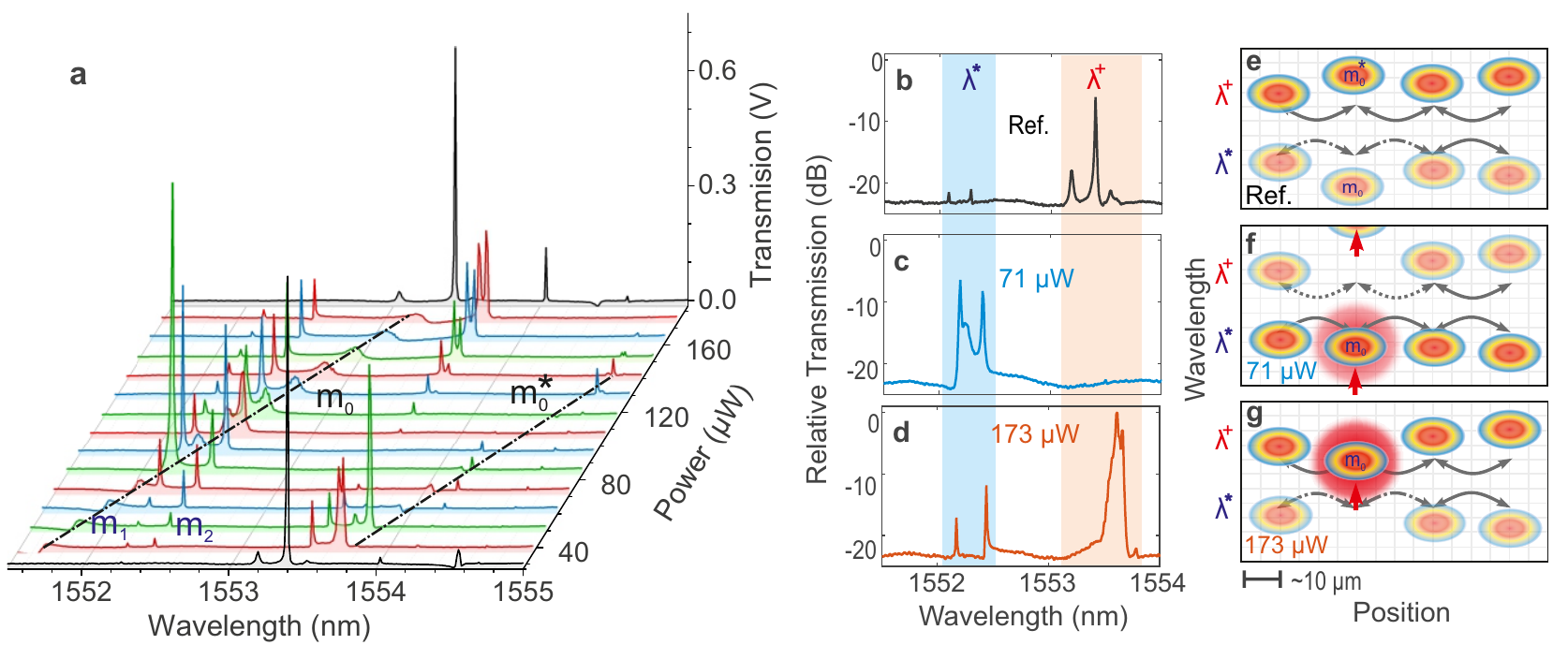}
\caption {\emph {
Measured and schematic representation of the localized modes that form a necklace state in a photonic crystal waveguide and the control of the necklace links.
(a) The transmission through the GaInP photonic crystal waveguide illustrating the full range localized modes while the power on the linking modes are scanned between \uW{37} and \uW{173}.
The dashed dotted black lines trace the resonance wavelength of the link modes $m_{0}$ and $m_{0}^*$. The first and the last black curves are the reference measurements before and after controlling the necklace states, respectively.
(b,e) The unperturbed localized modes in a photonic crystal waveguide.
At $\lambda^{*}$ light is weakly transmitted due to small spectral overlap of the modes, whereas at $\lambda^{+}$ light is transmitted given a larger spectral overlap of the mode $m_{0}^{*}$ with the rest of the necklace.
(c,f) The mode $m_{0}$ is locally tuned and thereby shifted in wavelength, which increases its coupling to the rest of the necklace modes at $\lambda^{*}$. 
Thereby the transmission is increased. The necklace link at $\lambda^{+}$ is incomplete given that the mode $m_{0}^{*}$ is shifted away.
(d,g) The pump power applied to the link mode $m_{0}$ is increased to \uW{173} which further shifts the $m_{0}$ mode and it now completes the necklace at $\lambda^{+}$.
In panels e-g, the dimmed modes and dotted lines represent low transmission whereas bright colored modes and solid lines show increased transmission. See~Supplementary Fig.\ref{FigSup1} for additional data that provide proof of robustness of our reversible control.
}}
\label {Fig2}
\end {center}
\end {figure*}

Figure~\ref{Fig2}(a) shows transmission through our GaInP photonic crystal waveguide versus wavelength at various pump powers.
One can observe two small transmission features, corresponding to modes labeled $m_0$ and $m_0^*$, that shift strongly with pump power.
The other transmission features do not shift appreciably, which indicates that we have independent control over the resonance frequencies of localized modes (see Fig. \ref{Fig2}) thanks to the narrow spatial profile of our pump. 
The pump beam has a spot size of \um{0.96} and the resulting temperature profile extends up to \um{5} \cite{sokolov.2015.apl}.
A group of modes  between \nm{1552} and \nm{1552.5} is not shifted in frequency, indicating negligible spatial overlap with the pump. 
However, the transmission of these modes is strongly increased when they become resonant with $m_0$.  
Similarly,the transmission of the modes between \nm{1553.25} and \nm{1553.75} is decreased, as they lose spectral overlap with mode $m_0^*$ and recovers as they regain spectral overlap with mode $m_0$. The pump beam is kept on  modes $m_0$ and  $m_0^*$ during all the measurements.
Before and after the tuning experiments we obtain reference spectra, shown as the very first and the last curves, to validate that there is no permanent change on the sample. 
Indeed the second reference spectrum is almost identical a small decrease in intensity attributed to drift of coupling losses that only affect the total transmission and not the spectral properties.

In Fig.~\ref{Fig2}(b-d) we present salient features of the transmission spectra at a higher resolution and in dB scale which shows the modulation depth more clearly. 
Fig.~\ref{Fig2}(b) depicts a reference transmission spectrum. 
In the reference spectrum we detect weak transmission peaks in the wavelength region marked by $\lambda^{*}$ (\nm{1552} to \nm{1552.5}) and stronger transmission in the region $\lambda^{+}$ (\nm{1553.25} to \nm{1553.75}). 
Next, we shine \uW{71} of pump laser light on the mode $m_0$, which makes it resonant with the modes located at $\lambda^{*}$ region.  As a result transmission at $\lambda^{*}$ is increased by 15 dB and  transmission at $\lambda^{+}$ is decreased to the noise level, see Fig.~\ref{Fig2}(c). 
Finally, the power on the $m_0$ mode is increased to \uW{173}, which brings $m_0$ into resonance with $\lambda^{+}$, and consequently the transmission  at $\lambda^{+}$ increases by 23 dB, while the transmission at $\lambda^{*}$ decreases, see Fig.~\ref{Fig2}(d). 
While we are tuning $m_0$ mode away from $\lambda^{*}$ region, the spectral overlap of mode $m_0$, given the broad Fano profile, is still maintained up to a certain degree. 
Moreover, our method enables us to mark only the localized modes that we can identify in transmission. 
The system has more localized modes that are out of resonance and we may not be able to resolve a transmission peak for these modes. 
As we tune a localized mode, more modes can possibly couple to the necklace and contribute to the transmission. 
For these reason, the signal at  $\lambda^{*}$ region does not drop to noise level as in $\lambda^{+}$ region.
This strong modulation of the transmission shows that we can control necklace states by independently controlling one of the necklace links.
 
A schematic model of these necklace states is shown in Fig.~\ref{Fig2}(e-g).
In Fig.~\ref{Fig2}(e) we depict our interpretation of the reference spectrum: In this case the link mode $m_0$ is out of resonance leading to low transmission. A spatially nearby $m_0^*$ mode enables a weak transmission at $\lambda^+$. 
When we tune the refractive index locally, the $m_0$ mode is tuned into resonance and completes the chain at $\lambda^{*}$, leading to an increased transmission. 
At the same time, the mode $m_0^*$, is pushed out of resonance with the necklace state at $\lambda^{+}$, decreasing transmission in that part of the spectrum.
In the third step (Fig.~\ref{Fig2} d,g) the power on the $m_0$ mode is increased to \uW{173}, which induces a greater shift in wavelength.
At this power level the $m_0$ mode decouples from the necklace state at $\lambda^{*}$ and completes the chain at $\lambda^{+}$.
As a result, the transmission at $\lambda^{*}$ is decreased, while the chain at $\lambda^{+}$ shows a high transmission. 
The independent control of one of the link modes in a necklace state enables us to switch the wavelength at which the sample becomes transmitting.

\begin {figure*}[htb]
\begin {center}
\includegraphics {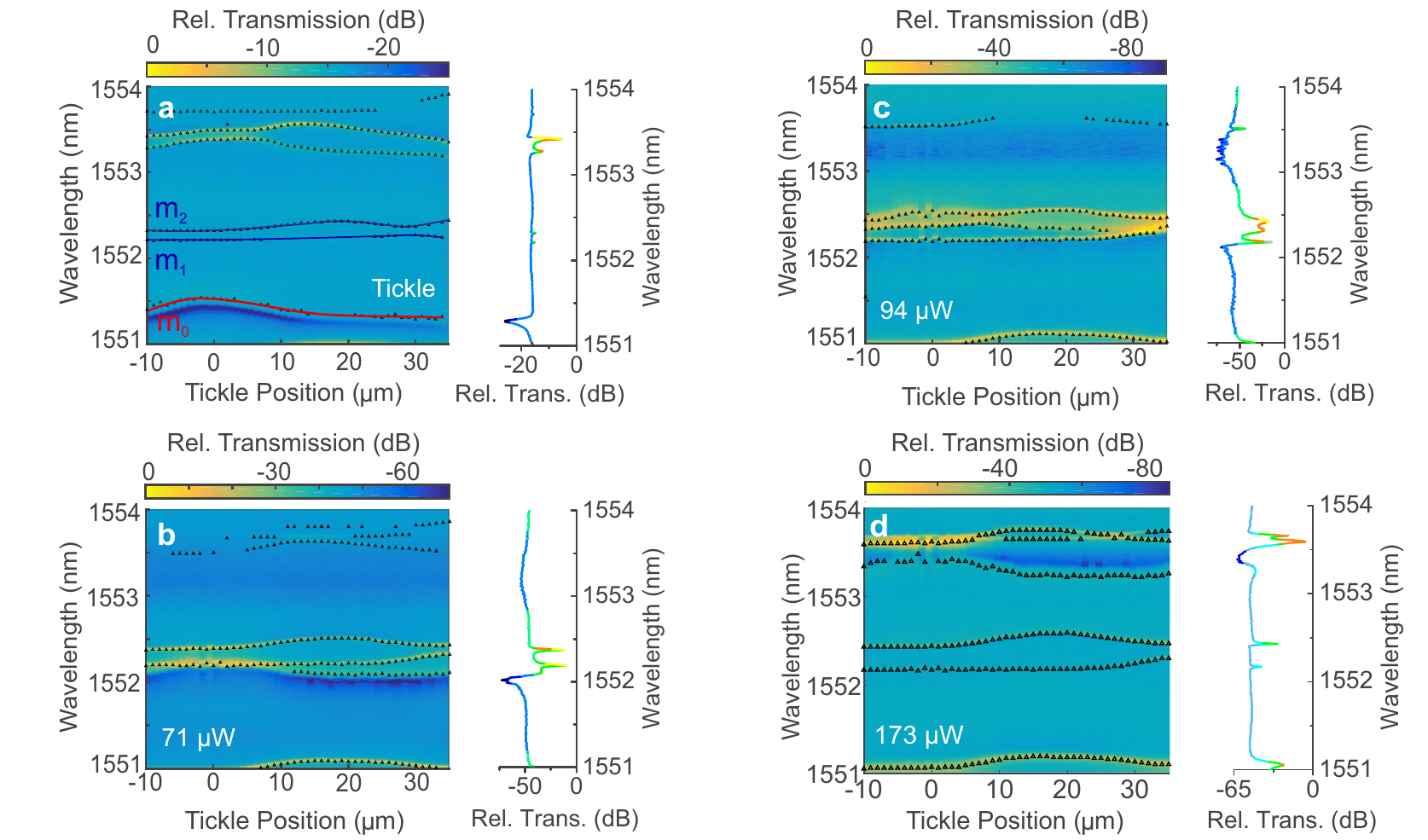}
\caption {\emph {
Control of the necklace states in a photonic crystal waveguide, mapped in wavelength and in position.
The density plots show transmission versus wavelength and position of the tickle beam.
The graphs at the side of each panel show relative transmission versus wavelength, taken at a tickle spot position at \um{35}.
(a) The $m_{0}$ mode is located spatially using a tickle beam. 
The modes $m_1$ and $m_2$ are located at \nm{1552.2} and \nm{1552.4}, respectively.
The mode at $m_1$ is spatially centered at \um{29} and $m_2$ is located at \um{18}.
The pump beam is positioned at 0, on top of $m_{0}$. In panels (b-d) the power is increased step-wise. 
Panel (b): at \uW{71}, $m_{0}$ weakly couples to $m_1$. 
Panel (c): strong coupling of  $m_{0}$ to $m_2$ 
(d) The $m_{0}$ mode is decoupled from the necklace at \uW{173} pump power. 
The black and red solid bold curves are guides to the eye. see~Supplementary Fig.\ref{FigSup3} for intermediate power steps.
}}
\label {Fig3}
\end {center}
\end {figure*}

Detailed spatial information on the coupled modes is represented in Fig.~\ref{Fig3}, showing a color map of the transmission versus frequency and tickle beam position. In Fig.~\ref{Fig3}(a) only the tickle beam is present. The tickle beam induces a small index perturbation and that effectively shifts the mode by about \nm{0.24}.
From the maxima of the tickle-induced wavelength shifts we locate mode $m_0$ at \um{0}, mode $m_1$ at \um{29}, and mode $m_2$ at  \um{18}. 
From the spectrum in  Fig.~\ref{Fig3}(a) we see that the line shape of  mode $m_0$ resembles a Fano profile, which arises from the interference of the discrete (localized) resonance with a transmission continuum~\cite{fano.1961.pr, Galli.2009.apl, kivshar.2010.rmp, fan.2012.apl, pfeifer.2013.science, lian.2016.pra}, see Supplementary Fig.~\ref{FigSup2}.
Next, we position the pump beam at position 0, to spatially overlap with $m_0$, and tune the refractive index locally. 
The hybridization and the anti-crossings of the localized modes \cite{Bliokh.2008.prl} are observed at the position of the linking mode. 
In Fig.~\ref{Fig3}(b) we see that mode $m_0$ shifts in frequency by \nm{0.9}, due to the pump and the tickle beam, and overlaps in frequency with $m_1$.
We increase the pump power to \uW{94} as shown in Fig.~\ref{Fig3}(c). 
In this case, mode $m_0$ couples to mode $m_2$ strongly as is apparent from the wide avoided crossing.
Scanning the tickle beam on top of a pump provides us the means to measure the anti-crossing width.
Although the tickle beam induces a wavelength shift of \nm{0.24} typically, we observe that when mode $m_0$ is at the vicinity of mode $m_2$ the shift induced by the tickle beam is much smaller (\nm{0.05}), as expected for coupled modes in an anticrossing.
In Fig.~\ref{Fig3}(c)  the coupled modes obtain a flat spatial profile which indicates the increased spatial size of the  hybridized modes.
When two modes weakly couple the spatial mode profile get broader in space due to hybridization (crossing of the imaginary part of the eigenvalues). 
When the modes are strongly coupled (anti-crossing of the real part of the eigenvalues), the modes repel each other, which results is in shifting of the position as well as resulting in weaker or stronger transmission as can be seen in Fig.~\ref{Fig3}.
In supplementary Fig.~\ref{FigSup3}, we provide measurements at intermediate pump power levels.
Finally, the pump power is increased to \uW{173} on the mode $m_0$. At this  power level the $m_0$ mode is decoupled again, it is now at higher wavelength than  modes $m_1$ and $m_2$.

The pump-dependent hybridization of modes  demonstrates that we can locally tune a localized mode and couple it to other localized modes weakly or strongly. 
The coupling strength is determined by the spatial position of the modes.
For instance, when we tune mode $m_0$ to the same frequency as mode $m_1$, we observe no avoided crossings  since the distance between these modes is as large as \um{29}, making them weakly coupled.
However, between modes $m_0$ and $m_2$, which are close to each other in space, a wide avoided crossing of \nm{0.1} is measured.

Using spatial control of the intrinsically localized modes in a waveguide we form, or break, highly transmissive necklace states so that the transmission can be modulated at a given wavelength.
The pump-tickle-probe method that we introduce here provide us the means to identify the coupling type, trace the anti-crossing regime, and detect spatial mode profile changes of a collection of disordered high-Q localized modes and extended necklace states. 
In more complex geometries, controlling multiple localized modes in a necklace state via holographic patterns can enable to route light on chip in 2D optical networks by coherently coupling Anderson-localized modes \cite{lodahl.2010.science, Ren-Jye.2017.Nanophotonics}.
This control of coupled narrow-band resonant modes is an essential step in the coupling of quantum light sources that can be embedded inside photonic crystal waveguides, and which offer novel opportunities for creating multi-node quantum networks~\cite{greentre.2006.natphys,kimble.2008.nature, Stockill.2017.prl, Lodahl.2018.QuanSciTEch}.

\bibliographystyle{naturemag}
\bibliography{C:/Users/eyuce/Dropbox/Library/151115_references}

\newpage


\medskip
\textbf{Acknowledgements} We thank Willem L. Vos, Ad Lagendijk, Pepijn W. H. Pinkse, Henri Thyrrestrup, and Sanli Faez for useful discussions. This research was supported by ERC-pharos. 




\end{document}